\documentclass[aps,prb,twocolumn,floatfix,showpacs]{revtex4-1}
\usepackage{graphics}
\usepackage{graphicx}
\usepackage{color}
\usepackage{amssymb}
\usepackage{amsmath}
\usepackage{subfigure}

\begin{document}
\title{Qubit state detection using the quantum Duffing oscillator}
\author{V. Leyton$^1$,  M. Thorwart$^1$, and V. Peano$^2$}
\affiliation{   $^1$I.\ Institut f\"ur Theoretische Physik, Universit\"at
                    Hamburg, Jungiusstra{\ss}e 9, 20355 Hamburg, Germany \\
                $^2$Department of Physics and Astronomy, Michigan State
                    University, East Lansing, MI 48824, USA}
\date{\today}
\begin{abstract}
We introduce a detection scheme for the state of a qubit, which is based on
resonant few-photon transitions in a driven nonlinear resonator.  The latter is
parametrically coupled to the qubit and is used as its detector.  Close to the
fundamental resonator frequency, the nonlinear resonator shows sharp resonant
few-photon transitions. Depending on the qubit state, these few-photon
resonances are shifted to different driving frequencies.  We show that this
detection scheme offers the advantage of small back action,  a large
discrimination power with an enhanced read-out fidelity, and a sufficiently
large measurement efficiency. A realization of this scheme in the form of a
persistent current qubit inductively coupled to a driven SQUID detector in its
nonlinear regime is discussed. 
\end{abstract}
   
\maketitle

\section{Introduction}

The efficient and reliable detection of the quantum mechanical state of a
nanoscale system is a key component of all present designs of quantum  
circuits.\cite{Clerk10} One nondestructive readout scheme currently in use for
the important class of superconducting flux qubits is based on a heterodyne
detection of the dynamic response of a dc superconducting quantum interference
device (dc-SQUID) detector which is inductively coupled to the qubit.
\cite{Lupascu04,Lupascu05} Thereby, the dc-SQUID is operated in its linear
regime as a shunted variable inductor in a resonant circuit. In this set-up,
 its resonance frequency depends on the magnetic flux generated by the
qubit being in the ground or excited state. Hence, measuring  
the impedance of the resonant circuit as a function of an externally 
applied bias current yields two characteristic Lorentzian resonances
 at two different resonance frequencies, which depend on 
the two qubit states. This detection scheme, hence, allows us to infer the 
state of the qubit from the resonant response of the detector in
the nanocircuit. In order that a reliable discrimination of the two qubit
states becomes possible in this continuous type of readout design, the
probability distributions for the readout values have to be only weakly
overlapping. Due to thermal and quantum fluctuations, the readout 
naturally is a random process,\cite{Lupascu05} and the noise properties of the
nanocircuit around the detector resonances determine the discrimination power of
the set-up.  

An alternative readout scheme is the Josephson bifurcation
amplifier.\cite{Siddiqi04,Vijay09}  It is based on a classical driven nonlinear
resonator and exploits the classical bifurcation point of the dynamically
induced bistability with a small- and a large-oscillation state.\cite{Nayfeh79}
The response (or output) of the nonlinear resonator around the bifurcation
point is very sensitive to small changes in the circuit parameters. This is an
ideal prerequisite for a sensitive detector. 
Depending on the state of the qubit to be sensed, the resonator
bifurcation point is shifted to a different frequency, allowing for large
discrimination powers between the large- and small-oscillation detector state of
up to $98 \%$.\cite{Lupascu06} Nevertheless, since the detector is a
classical macroscopic device, it introduces considerable dephasing and
relaxation to the qubit state, yielding a
reduced contrast of the qubit Rabi oscillations of less than $90
\%$.\cite{Lupascu06} This implies that the thermal noise properties of 
the nonlinear detector (together with semiclassical corrections due to quantum
fluctuations) around the classical bifurcation point determine the
discrimination power between the two states close to the classical bifurcation
point. \cite{Dykman79,Dykman90,Dmitriev86,Dykman88} Hence, it would be
desirable to combine the advantage of a large discrimination power of a
nonlinear detector with the reduced noise sensitivity of a nanocircuit operated
close to the quantum regime. 

In this paper, we introduce a combination of both strategies and 
propose a nonlinear detector scheme in the form of a nonlinear resonator with
an amplitude modulated drive in its few-photon deep quantum regime. In
particular, in this regime, we shall exploit sharp multiphoton resonances in the
nonlinear resonator, \cite{Peano04,Peano06a,Peano06b} which are induced by the
external driving field close to the fundamental resonator frequency. They can be
used for the detection of the states of the 
qubit and offer the advantage of being rather sharp and externally tunable by
varying the parameters of the external drive. The concept is an extension
of the case of a linear
resonator, where the fundamental resonance frequency is shifted depending on
the qubit state. However, the multiphoton resonances in the nonlinear
detector close to the  detector's fundamental frequency show 
very small line widths. The width of the $N$-photon resonance is determined by
the corresponding $N$-photon Rabi frequency, which decreases with increasing 
photon number. The sharp resonance lines, in turn, offer the advantage that
only a few measurement cycles are necessary to ensure a large discrimination
power. To understand the back action of the nonlinear multiphoton
detector on the qubit state,  we determine the relaxation rate of the qubit
due to the coupling to the driven dissipative nonlinear oscillator around a
multiphoton resonance. Notably, the back action of the resonator on
the qubit is sufficiently weak, yielding to a good qubit-state
measurement fidelity.
Furthermore, we show that the discrimination power of the set-up is rather
large and beyond $98 \%$ for our choice of realistic parameters of a flux
qubit circuit. In fact, it gives rise to an enhanced measurement 
fidelity as compared to the linear parametric oscillator. 
Furthermore, we show that the nonlinear multiphoton detector does not have 
a worse measurement efficiency as compared to the linear detector
scheme. We determine the measurement efficiency of the set-up via the
ratio of the time it takes to collect enough information on the qubit state
(measurement time) and the relaxation time. It turns out that the measurement
efficiency does not considerably decrease as compared to the linear case. Hence,
the detection scheme indeed has the advantage of an overall reduced back action
in combination with an enhanced discrimination power, together with a
sufficiently large measurement efficiency. 

An experimental realization of a driven nonlinear resonator in its
few-photon quantum regime is in principle possible with present set-ups and
technology. In a recent experiment,\cite{Murch10} a nanoscale superconducting
microwave resonator has been driven to its nonlinear regime by fast
frequency-chirped voltage pulses. At low enough temperature, the regime of
quantum noise has been reached. In this experiment, the applied driving strength
has been rather large, which corresponds to a large photon number transferred to
the resonator. No particular few-photon resonances have been
revealed and the nonlinear response is similar to previous schemes on
classical bifurcation detectors using a time-dependent driving
frequency.\cite{Naaman08}. However, the route to 
the few-photon regime seems to be clear. 

The paper is organized as follows. In Sec.\ \ref{sec:Model}, we start from a
typical experimental setup for the flux qubit and its SQUID detector and we
derive the Hamiltonian model. This serves to motivate an experimental 
realization of our proposed detection scheme. Moreover, we discuss the regime of
validity of the model.  In accordance with the approximation made in Sec.\
\ref{sec:Model}, we
continue the study of the coherent dynamics in Sec.\ \ref{sec:Floquet} in the
rotating
wave approximation. Dissipative coupling to the environment is included on the
level of a Born-Markov master equation in the rotating frame in Sec.\
\ref{sec:Diss}. In Sec.\ \ref{sec:DetDyn}, we analyze the multiphoton
transitions in the nonlinear response of the Duffing oscillator and show that
their resonance frequency depends on the qubit state.  Then, in Sec.\
\ref{sec:BackQUBIT} we determine the back action of the driven dissipative 
detector on the qubit dynamics by analyzing the population difference
of the qubit states at the multiphoton transitions in the detector. 
Furthermore, we determine the measurement efficiency in Sec.\ \ref{sec:queff}. 
\section{\label{sec:Model} Model}
In order to relate the theoretical approach in the following to realistic
devices, we start by deriving the model from standard set-ups already realized
in experiments.  For this, we use a typical architecture of a persistent 
current qubit which is inductively coupled to a driven SQUID. 

\subsection{Persistent current qubit}
We consider the experimental set-up used in Ref.\ \onlinecite{flux-qubit} for
the qubit, consisting of a superconducting loop interrupted by three Josephson
junctions, two of which have equal Josephson energies, while the coupling
energy of the third is smaller, in order to yield a double-well potential
configuration. In this low-inductance circuit, the flux through the loop remains
close to the externally applied value $\Phi_{\rm  qb}$. When the latter is
close to $(n + 1/2)\Phi_0$, where $n$ is an integer and $\Phi_0$ is the flux
quantum, the device is described by the Hamiltonian in terms of the Pauli
matrices $\sigma_{x,z}$ as ($\hbar=1$)
\begin{equation}\label{qubitModel}
 H_{\rm  qb} = \frac{\epsilon}{2} \,   \sigma_{z} - \frac{\Delta}{2}  \,
 \sigma_{x} ,
\end{equation}
with the two eigenstates $|\!\!\uparrow \rangle $ and $|\!\!\downarrow \rangle $
of 
$\sigma_{z}$ corresponding to the two persistent current states 
 $\pm I_p$. The minimal energy level splitting
$\Delta$ and the
current $I_p$ are determined by the charging and Josephson energies of the
Josephson junctions. The asymmetry is given by $\epsilon = 2 I_p (\Phi_{\rm 
qb}-\Phi_{\rm  0} / 2)$. In the energy eigenbasis, the Hamiltonian follows as
$H_{\rm  qb} = \omega_{\rm  qb} \tau_z /2$,  with $\omega_{\rm  qb} =
\sqrt{\epsilon^2 + \Delta^2}$, and $\tau_z = \sigma_z \cos \theta - \sigma_x
\sin \theta$ is the corresponding Pauli matrix with $\tan\theta =
\Delta/\epsilon$. The detection of the qubit state essentially involves the
measurement of the magnetic flux produced by the persistent current states. To
this end, one can use the driven SQUID as a sensitive magnetometer,
\cite{Lupascu04} operating in its nonlinear region. Below, we will restrict to
the few-photon deep quantum regime. 
\subsection{Driven SQUID as a nonlinear quantum detector}
We consider the standard setup of a dc-SQUID formed by two Josephson junctions
in a superconducting loop, but subject to a time-dependent external bias 
current.\cite{Vijay09} Moreover, we assume a negligible ring inductance $L_R$ of
the SQUID (low-inductance approximation). \cite{Soerensen76}
In this configuration, the superconducting phase
differences at each junction, $\chi_{1}$ and $\chi_2$, play the role of
dynamical variables with a constraint given by the flux quantization, i.e.,
$\chi_1 - \chi_2 = - \Phi_{\rm sq}/\varphi_0 \equiv -2 \pi \varphi_{\rm  ex}$,
where $\Phi_{\rm  sq}$ is the external magnetic flux piercing the
superconducting loop and $\varphi_0 = \Phi_0 / 2\pi $. Note that 
within the low-inductance approximation, $L_R I_{\rm  0c} \ll \varphi_0$ with
the critical current $I_{\rm 0c}$ of the SQUID.  Thus, the
system is described by the
generalized coordinate $\chi_+ = (\chi_1 + \chi_2)/2$, with the effective
Lagrangian \cite{Clarke01}
\begin{eqnarray}
\label{lagrangian}
 L_{\rm  sq}(\chi_+, \dot{\chi}_+,t) &=&  \varphi_0^2 \, C_{0} \, 
\dot{\chi}_{+}^2 + E_J \cos\left(\pi \varphi_{\rm  ex}\right)\cos(\chi_{+})
\nonumber \\ 
&&  -\varphi_0  I_{\rm  b}(t) \, \chi_+,
\end{eqnarray}
where we have assumed a symmetric loop, with $E_J= \varphi_0 I_{\rm  0c}$ as the
Josephson energy, and $C_0$ as the capacitance of each junction. Moreover, we
include a time-periodic ac current
$I_{\rm  b}(t)=I_0 \cos (\omega_{\rm  ex} t)$ with frequency $\omega_{\rm ex}$
and amplitude $I_0$ injected ``into'' the loop. The above Lagrangian describes
an effective superconducting loop (with a negligible ring inductance) with a
single Josephson junction
\cite{Lupascu05} with a tunable Josephson energy $E_J \cos (\pi \varphi_{\rm 
ex})$, critical current $I_{\rm  c}=2I_{\rm  0c}|\cos \pi \varphi_{\rm   ex}|$,
cross-junction phase difference $\chi_+$, and capacitance $C=2C_0$.  In order to
tune the resonance frequency, the  SQUID is shunted \cite{Vijay09} with a
capacitance $C_{s} \gg C$. Next, we shall establish the optimal working point of
the qubit-detector system, where the dissipative influence entering via the
detector is minimal. 

\subsubsection{Qubit-detector interaction}
The qubit and the SQUID are coupled by means of their mutual inductance
$M$.\cite{Lupascu05,Makhlin01}  Thereby, the SQUID induces the
flux $M I_{\circlearrowleft}$ in the qubit loop, where $I_{\circlearrowleft}$ is
the circulating current in the SQUID. The latter can be determined by using
current conservation in the loop and the Josephson relations for the two
junctions in the SQUID. For the symmetric SQUID,\cite{Lupascu05} it follows that
$I_{\circlearrowleft}(t)=I_{\rm  c0} \sin (\pi \varphi_{\rm  ex}) \cos
(\chi_+(t))$. Thus, the total magnetic flux in the qubit is affected by
its coupling with the SQUID, and it is composed of the external flux and the
SQUID-generated contribution, i.e., $\Phi_{\rm  qb} \rightarrow \Phi_{\rm  qb} +
M I_{\circlearrowleft}(t)$. This implies that the energy bias of the qubit
acquires a contribution that depends on the circulating current in the SQUID,
leading to the effective asymmetry $\epsilon_f= \epsilon (\Phi_{\rm  qb}) +
\beta(I_{\circlearrowleft}(t))$, where $\beta(I_{\circlearrowleft}(t))= 2M
I_pI_{\circlearrowleft}(t)$.

Therefore, two sources of noise can affect the qubit dynamics, i.e.,  the
fluctuations from the external flux $\Phi_{\rm  qb}$ and from the bias current
$I_{\rm  b}(t)$ in the SQUID,\cite{Bertet05} which is related to $\chi_+$ by
the Josephson equation $I_b(t) = I_{c0} \sin(\chi_+(t))$. By tuning the bias
current to the critical value $I^*_{\rm b}$ characterized by $(d\beta/dI_{\rm 
b} )_{I_{\rm b}=I^*_{\rm b}}=0$, the influence from current fluctuations in the
SQUID can be minimized  \cite{Bertet05} and the optimal working point is
reached. For a nonsymmetric SQUID, the lowest-order contribution is linear in
$I_b$,\cite{Clarke01,Bertet05} while in the symmetric case this lowest-order
contribution vanishes, which implies that around the optimal working point the
phase $\chi_+$ is very small, $\chi_+ \sim 0$. In the following, we consider a
setup close to the optimal point, where we can expand the expression for
$I_{\circlearrowleft}$ up to second order in $\chi_+$, yielding the interaction
term 
\begin{equation}
  H_{\rm  qb-sq} = \tilde{g} \chi_+^2 \hat{\sigma}_z,
\end{equation}
with the coupling constant $\tilde{g}=2I_p I_{\rm  0c} M \sin (\pi \varphi_{\rm
ex})$.

\subsubsection{SQUID modelled as a Duffing oscillator}
As we operate the detector in its nonlinear regime, we expand the 
potential term $V(\chi_{+})= - E_J \cos\left(\pi \varphi_{\rm
ex}\right)\cos(\chi_{+}) \simeq V_0 + m \Omega^2 \chi_{+}^2/2 - \tilde{\alpha} 
\chi_{+}^4$ in Eq. (\ref{lagrangian}) around the optimal point up to fourth
order in $\chi_+$, where  $m= \varphi_0^2 C_{\rm s}$ is the effective mass,
$\Omega = \left( I_{\rm c} / \varphi_0 C_{\rm s} \right)^{1/2}$ the
corresponding frequency,  and $\tilde{\alpha} = m \Omega^2/4$ the strength of
the nonlinearity.  We switch to a description in terms of the creation and
annihilation operators $a$ and $a^\dagger$, defined by $\chi_+ = \chi_0 (a +
a^\dagger)$ with $\chi_0 = \sqrt{1/(2m\Omega)}$ the zero-point energy of the
phase $\chi_+$.  Adding the time-dependent driving term yields us to the
Hamiltonian of the driven SQUID described by the quantum Duffing oscillator
model 
\begin{equation}
H_{\rm  sq} = \Omega \, a^\dagger a-\frac{\alpha}{12} (a+a^\dagger)^{4}+
f(a+a^\dagger) \cos \left(\omega_{ \rm ex}t\right),
\end{equation}
with nonlinearity and driving strength given by $\alpha = 3I_c \varphi_0
\chi_0^4$, and $f = I_0 \varphi_0 \chi_0$, respectively.  Similarly, the
interaction Hamiltonian in terms of ladder operators reads as
\begin{equation} \label{squidH}
H_{\rm  qb-sq}  = \frac{g}{2} (a+a^{\dagger})^{2} \sigma_z,
\end{equation}
with $g=2 \tilde{g} \chi_0^2$.

\begin{figure}[t]
\includegraphics[width=70mm]{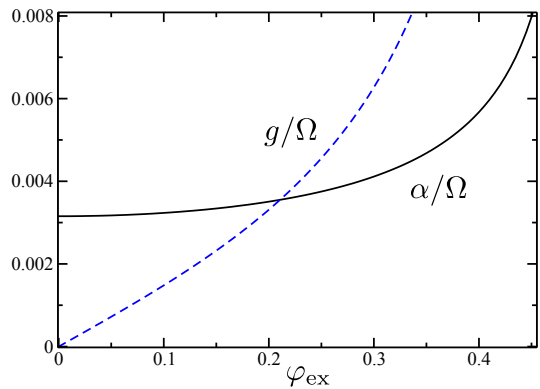}
  \caption{(Color online) Dependence of the dimensionless ratios
           $\alpha/\Omega$ and $g/\Omega$ on the external flux $\varphi_{\rm
           ex}$ in the SQUID. The parameters of the SQUID are chosen as $C_s =
           7.65$ pF, $I_{c0} = 200$ nA, $I_p = 300$ nA, and $M = 40$ pH.
           \cite{Lupascu05} \label{fig0}}
\end{figure}

Notice that $g$ and $\alpha$ depend on the external flux $\varphi_{\rm  ex}$,
i.e., they are tunable in a limited regime with respect to the desired
oscillator frequency $\Omega$, where the coupling term is considered as a
perturbation to the SQUID ($g<\alpha$), in order to keep the dynamics of the
oscillator to dominate. The dependence of the dimensionless ratios
$\alpha/\Omega$ and $g/\Omega$ is shown in Fig.\ \ref{fig0}. 
We restrict to parameters of the external magnetic flux in the SQUID loop, which
generate a weak nonlinearity and a weak qubit-detector coupling strength,
$\{\alpha, g\} \ll \Omega$, i.e., for $\varphi_{\rm ex} \sim 0$.  A typical
dependence of both parameters for typical experimental parameters is shown in
Fig.\ \ref{fig0}. Both cases of $g/\Omega > \alpha/\Omega$ and $g/\Omega <
\alpha/\Omega$ can be achieved. For our purpose of a qubit-detector setup, the
qubit-resonator coupling typically will be required as small enough in order to
ensure a minimal back action. On the other hand, the qubit-detector coupling
should be large enough so that an efficient detection of the qubit state becomes
possible.  As is shown in Fig.\ \ref{fig0} and will be quantitatively discussed
in the sequel of this paper, this can indeed be achieved for realistic
parameters. Moreover, the choice of the parameter regime also justifies us to
restrict the influence of the resonator coupling on the effective qubit bias to
lowest order in $\chi_+$ only.  Eventually, the total system is described by the
Hamiltonian $H(t)= H_{\rm qb} + H_{\rm  qb-sq} + H_{\rm sq}(t)$.
\section{\label{sec:Floquet} Coherent dynamics and rotating-wave approximation
}
Before we address the dynamics of the detection scheme based on the nonlinear 
response of the Duffing oscillator to the applied periodic driving in the
stationary regime, we discuss the coherent dynamics generated by $H(t)$,  which
is periodic in time. 

Here, we are interested in exploiting few-photon transitions in the detector
around the fundamental detector frequency $\Omega$. Hence, higher harmonics have
a small amplitude and can effectively be neglected.  Furthermore, we focus on
the regime of weak nonlinearity, weak driving, and weak qubit-detector coupling
as characterized by $\lbrace \alpha,f,g\rbrace \ll \Omega$.  The proposed
mechanism of detection is most conveniently discussed in the simplest case, when
the dynamics occurs close to the fundamental oscillator resonance $\omega_{\rm 
ex} \sim \Omega \sim \omega_{\rm qb}/2$. Then, the rotating-wave approximation
(RWA) can be invoked in order to obtain a simple interpretation in terms of
few-photon transitions. In passing, we note that we have also performed a
complete analysis in terms of full Floquet theory, thereby avoiding the RWA. For
all cases shown below, both approaches yield coinciding results.

We switch to the rotating reference frame by the transformation $R(t)=\exp\{
i(a^{\dagger}a+\tau_z)\omega_{\rm  ex}t \}$. Then, the RWA eliminates the fast
oscillating terms from the transformed Hamiltonian ${\cal
H}=R(t)H(t)R^\dagger(t)-i R(t)\dot{R}^\dagger (t)$ and the time-independent
Schr\"odinger equation in the rotating frame ${\cal H} |\varphi_\alpha\rangle
=\varepsilon_\alpha |\varphi_\alpha\rangle$ follows, with the RWA Hamiltonian
given by

\begin{equation}\label{RWAhamiltonian}
{\cal H} = {\cal H}_{\rm qb} + {\cal H}_{\rm qb-sq} + {\cal H}_{\rm sq} \, ,
\end{equation}
with
\begin{eqnarray}
&&{\cal H}_{\rm qb} =   \frac{1}{2} \delta \omega_{\rm  qb} \tau_z , 
\nonumber \\
&&{\cal H}_{\rm qb-sq} = g\cos \theta \, a^\dagger a \, \tau_z+ \frac{g}{2}\sin
\theta \, (a^{\dagger\, 2} \tau^- + a^2 \tau^+ ) \, , 
\nonumber \\ 
&&{\cal H}_{\rm sq} =  \delta \Omega \, a^\dagger a - \frac{\alpha}{2} a^\dagger
a
\, a \, a^\dagger +\frac{f}{2} (a +a^\dagger) \, .  \nonumber
\end{eqnarray}
The detuning frequencies follow as $\delta \Omega=\Omega-\omega_{\rm  ex}$ and
$\delta \omega_{\rm qb} = \omega_{\rm  qb} - 2\omega_{\rm  ex}$, and $\tau^\pm =
(\tau_x \pm i \tau_y)/2$. The quasienergies $\varepsilon_\alpha$ and the RWA
eigenstates $| \varphi_\alpha \rangle$ result from a straightforward numerical
diagonalization of ${\cal H}$. In the static frame, an orthogonal (at equal
times) set $\{ |\tilde{\varphi}_{\alpha}(t)\rangle \}$ of approximated solution
of the Schr\"odinger equation follows as \begin{equation}\label{FloqueRWA}
|\tilde{\varphi}_{\alpha}(t) \rangle \simeq e^{-i\varepsilon_{\alpha} t}
|\phi_\alpha(t) \rangle = e^{-i\varepsilon_{\alpha} t}
e^{-i(a^{\dagger}a+\tau_z)\omega_{\rm  ex}t} | \varphi_\alpha \rangle.
\end{equation}
Here, the quasienergy states $|\phi_\alpha(t) \rangle  \equiv
e^{-i(a^{\dagger}a+\tau_z)\omega_{\rm ex}t} | \varphi_\alpha \rangle$
are time periodic with period $2\pi/\omega_{\rm ex}$ and form a complete basis
that will be used below for the description of the dissipative dynamics.  We
note that an analytic expression for the multi-photon resonances would follow
from a Van-Vleck perturbative approach in a similar manner as for the pure
quantum Duffing oscillator.\cite{Peano06a,Peano06b}  However, the resulting
expression will be cumbersome and not further illuminating for the present
purpose. We note, furthermore, that the qubit-detector interaction occurs via a
parametric coupling $g\cos \theta \, a^\dagger a \tau_z$, and via a two-photon
coupling $ g \sin \theta \, (a^{\dagger\, 2} \tau^- + a^2 \tau^+ )/2$. 
\section{\label{sec:Diss} Dissipative dynamics}
The electronic nanocircuit is embedded in a dissipative environment. In
particular, the SQUID is shunted with an Ohmic resistor, which yields
dissipative fluctuations $\xi(t)$.\cite{weiss}  We focus to the case of an
underdamped SQUID, where the shunt resistance is large, \cite{Makhlin01,Vijay09}
and use the standard harmonic bath in order to model the fluctuations, which are
rooted in current fluctuations and can be encoded in the Ohmic spectral density
$J(\omega)=\gamma \omega$.\cite{weiss} They couple to the resonator's dipole
operator, i.e., $H_{\xi} = \chi_+ \xi(t)$.  We note that, in the same way, the
direct coupling of the qubit to the electromagnetic fluctuations could be
included. However, we have checked \cite{Peano10} that for a related set-up of 
a flux qubit coupled to a harmonic oscillator, such a direct dissipation of the
qubit yields only minor quantitative corrections, which should be included in a
quantitative description of an experiment,\cite{Bertet05} but do not add
qualitatively new physics.  

The time evolution of the reduced density operator $\varrho(t)$ is described in
terms of a standard Markovian master equation projected onto the basis of 
the quasienergy states $\lbrace |\phi_\alpha(t)\rangle\rbrace$
\begin{equation} \label{FME}
\dot{\varrho}_{\alpha \beta}(t) =-i (\varepsilon_{\alpha} -
\varepsilon_{\beta})\varrho_{\alpha \beta}  + \sum_{\alpha' \beta'} {\cal
L}_{\alpha \beta, \alpha' \beta'} \varrho_{\alpha' \beta'}(t),
\end{equation}
where $\varrho_{\alpha\beta}(t) \equiv \langle \phi_\alpha(t) \lvert \rho(t)
\lvert \phi_\beta (t) \rangle$. The dissipative transition rates are given by
\cite{Peano06a,Peano06b} 
\begin{eqnarray}\label{Trates}
  {\cal L}_{\alpha \beta, \alpha' \beta'}  &=& 
  \sum_{n} (N_{\alpha \alpha',-n} + N_{\beta \beta',-n}) \chi_{\alpha
  \alpha',n}\chi_{\beta \beta',-n} \nonumber \\
  &&- \delta_{\alpha \alpha'} \sum_{\alpha'',n} N_{\alpha''\beta',-n}
  \chi_{\beta' \alpha'',-n} \chi_{\alpha'' \beta,n}
  \nonumber
  \\
  && - \delta_{\beta \beta'} \sum_{\beta'',n} N_{\beta''\alpha',-n}
  \chi_{\alpha \beta'',-n} \chi_{\beta'' \alpha',n}, \nonumber \\ &&
\end{eqnarray}
with $n\in \mathbb{Z}$ and $\chi_{\alpha\beta,n}$ being the Fourier components
according to $\langle \phi_\alpha(t) \lvert \chi_+ \lvert \phi_\beta (t)
\rangle=\sum_n \exp\lbrace -i\omega_{\rm ex} n t\rbrace\chi_{\alpha\beta,n}$.  
Furthermore, we have used the Planck numbers $N_{\alpha\beta,n} =
N(\varepsilon_\alpha -\varepsilon_\beta +n\omega_{\rm ex})$, where $
N(\varepsilon)= \gamma \varepsilon \left[\coth(\varepsilon/2T)-1 +
\Theta(-\varepsilon ) \right]$ with $k_B = 1$, temperature $T$ and $\Theta(x)$
being the Heaviside function. Since within the rotating wave
approximation, $ \lvert\phi_\alpha(t)\rangle \approx e^{-i(a^{\dagger}a+\tau_z)
\omega_{\rm  ex}t} | \varphi_\alpha \rangle$, the only non-zero Fourier
components are $\chi_{\alpha \beta,1} = \chi_0 \langle
\varphi_\alpha |a|\varphi_\beta \rangle / \sqrt{2}$\,, and $\chi_{\alpha
\beta,-1} = \chi_0 \langle \varphi_\alpha |a^\dagger|\varphi_\beta \rangle /
\sqrt{2}$ and the master equation (\ref{FME}) considerably simplifies as it
involves only single step transitions, i.e., one-photon emission (for $n=-1$)
into and absorption (for $n=+1$) processes from the bath. We note that
neglecting also the quasienergy dependence of the Planck numbers would yield the
well-known Lindblad master equation.

In order to measure the dynamic response of the resonator to the external drive
at asymptotically long times, a heterodyne detection scheme such as in Ref.\
\onlinecite{Bishop} can be used,\cite{Lupascu05} where the coupled
qubit-oscillator system approaches the steady state $\varrho^\infty=\varrho(t\to
\infty)$. We determine the stationary solution characterized by $\dot{\varrho}
(\infty)=0$ numerically. For this, we solve the corresponding eigenvalue problem
and $\varrho^\infty$ follows as eigenvector to the eigenvalue zero.  With this,
we compute the nonlinear response of the detector, characterized by the mean
value $\langle \chi_+ \rangle_\infty (t)$  at asymptotic times. As we restrict
the discussion to the regime close to the first harmonic (small detuning),
higher harmonics can be neglected and we immediately obtain   
\begin{eqnarray}
 \langle \chi_+ \rangle_\infty (t) & = & \mathrm{tr} (\varrho^\infty \chi_+)
\nonumber \\ 
& = & \sum_{\alpha,\beta}\rho^\infty_{\alpha\beta} \langle
\phi_\beta(t)|\chi_+|\phi_\alpha(t)\rangle 
\nonumber \\
& = &  \sum_{\alpha \beta} \varrho^\infty_{\alpha \beta} (\chi_{\beta
\alpha,+1} e^{i \omega_{\rm  ex} t} + \chi_{\beta \alpha,-1}  e^{ -i \omega_{\rm
 ex} t } ) \, . \nonumber \\
\end{eqnarray}
As the system is driven with frequency $\omega_{\rm  ex}$, $\langle \chi_+
\rangle_\infty (t)$ also oscillates with time. Its amplitude is given by 
\begin{equation}
A=  \sum_{\alpha\beta} \varrho_{\alpha\beta}
(\chi_{\beta\alpha,+1}+\chi_{\beta\alpha,-1}) \, .
\end{equation}
Correspondingly, we evaluate the population difference $\langle \sigma_z
\rangle_\infty(t)$ of the qubit states and obtain 
\begin{eqnarray}
\langle \sigma_z \rangle_\infty (t) &=& \mathrm{tr} (\varrho^\infty \sigma_z)
\nonumber \\ 
& = & \sum_{\alpha,\beta}\rho^\infty_{\alpha\beta} \langle 
\phi_\beta(t)|\sigma_z|\phi_\alpha(t)\rangle 
\nonumber \\
& = & \sin\theta \sum_{\alpha \beta}
\varrho^\infty_{\alpha \beta} (\tau^+_{\beta \alpha} e^{2i \omega_{\rm  ex} t} +
\tau^-_{\beta \alpha} e^{-2i \omega_{\rm  ex} t} ) 
\nonumber \\ 
&& + \cos\theta \sum_{\alpha \beta} \varrho^\infty_{\alpha \beta} \tau^z_{\beta
\alpha} ,
\end{eqnarray}
where $\tau^z_{\alpha\beta} =  \langle \varphi_{\alpha}| \tau_z
|\varphi_{\beta}\rangle$ and $\tau^\pm_{\alpha\beta}=  \langle \varphi_{\alpha}|
\tau^\pm |\varphi_{\beta}\rangle$. The population difference oscillates, with a 
maximal value given by 
\begin{equation}
P_\infty=\cos \theta \sum_{\alpha\beta} \varrho_{\alpha\beta} \,
\tau^z_{\beta\alpha}  + \sin \theta  \sum_{\alpha \beta}
\varrho_{\alpha\beta} \tau^x_{\beta\alpha} \, . 
\end{equation}
\section{\label{sec:DetDyn} Detector's dynamics}
\begin{figure}[t]
\includegraphics[width=80mm]{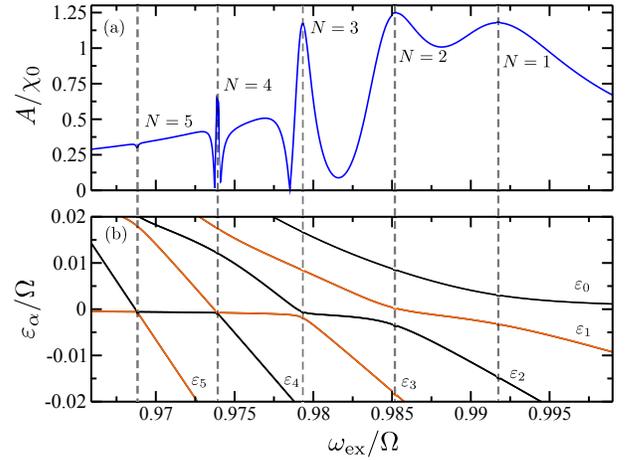}
  \caption{(Color online) (a) Amplitude $A$ of the nonlinear response of the
	   decoupled quantum Duffing 
           detector ($g=0$) as a function of the external driving frequency
           $\omega_{\rm  ex}$. (b) The corresponding quasienergy spectrum
           $\varepsilon_\alpha$. The labels $N$ denote the corresponding 
           $N$-photon (anti-)resonance. The parameters are $\alpha=0.01 \Omega$,
           $f=0.006 \Omega$, $T = 0.006 \Omega$, and $\gamma=1.6 \times 10^{-4}
           \Omega$. \label{fig1}}
\end{figure}
\subsection{No coupling between detector and qubit}
Before turning to the quantum detection scheme, we discuss the dynamical
properties of the isolated detector, which is the quantum Duffing oscillator. A
key property is its nonlinearity which generates multiphoton transitions at
frequencies $\omega_{\rm ex}$ close to the fundamental frequency $\Omega$. In
order to see this, one can consider
first the undriven nonlinear oscillator with $f=0$ and identify degenerate
states, such as $|n\rangle$ and $|N-n\rangle$ (for $N>n$), when $ \delta \Omega
= \alpha (N+1) / 2 $.\cite{Dykman05,Peano06a,Peano06b}  For finite driving $f >
0$, the degeneracy is lifted and avoided quasienergy level crossings form, which
is a signature of discrete multiphoton transitions in the detector.  As a
consequence, the amplitude $A$ of the nonlinear response signal exhibits peaks
and dips, which depend on whether a large or a small oscillation state is
predominantly populated.\cite{Peano06a, Peano06b}  The formation of peaks and
dips goes along with jumps in the phase of the oscillation, leading to
oscillations in or out of phase with the driving.  A typical example of the
nonlinear response of the quantum Duffing oscillator in the deep quantum regime
containing few-photon (anti-)resonances is shown in Fig.\ \ref{fig1}(a)
(decoupled from the qubit), together with the corresponding quasienergy
spectrum [Fig.\ \ref{fig1}(b)].  We show the multiphoton resonances up to a
photon number $N=5$. The resonances get sharper for increasing photon number,
since their widths are determined by the Rabi frequency, which is given by the
minimal splitting at the corresponding avoided quasienergy level crossing. 
Performing a perturbative treatment with respect to the driving strength $f$, 
one can get the minimal energy splitting at the avoided quasienergy level
crossing $(0,N)$ as \cite{larsen,Peano06a}
\begin{equation}\label{rabi}
 \Omega_{N,0} = f \left(\frac{2 f }{3 \alpha}\right)^{N-1}
\frac{\sqrt{(N)!}}{(N-1)!^2}.
\end{equation}
Because the nonlinearity $\alpha$ is typically fixed by the design of the
SQUID, the Rabi frequency can be easily tuned by tuning the driving strength
$f$.

\subsection{\label{nobackaction} Detector response for weak coupling to the
qubit}

Next, we consider a finite coupling of the detector to the qubit whose state is
to be sensed, i.e., $g \neq 0$. The coupling inevitably induces relaxation and
decoherence in the qubit, characterized by the relaxation and dephasing rate,
$\Gamma$ and $\Gamma_d$, respectively.  Typically, the detector couples only
weakly to the system, i.e., $g\ll \omega_{\rm qb}$. Then, the associated
relaxation and dephasing times ($T_1$ and $T_2$, respectively) are still much
larger than the corresponding relaxation time scale for the detector given by
$1/\gamma$. In passing, we note that the corresponding relaxation time around a
{\em resonant} multiphoton transition (in the underdamped case) has been shown
in Refs.\  \onlinecite{Peano04,Peano06b} to be comparable to $\gamma$. 
Moreover,
we  bias the qubit with a large asymmetry, $\epsilon\gg\Delta$ in order to 
``gauge'' the detector response.

For a rough evaluation of the order of magnitude of the involved time scales, we
may neglect the nonlinearity of the detector ($\alpha=0$) for the moment and
estimate the effective relaxation rate for the qubit coupled to an Ohmically
damped harmonic oscillator.\cite{Thorwart04}  This model can be mapped to a
qubit coupled to a structured harmonic environment with an effective
(dimensionless) coupling constant $\kappa_{\rm eff}=8 \gamma g^2/\Omega^2$.  For
the realistic parameters used in Fig.\ \ref{fig1} and $g=0.0012\Omega$, we find
that $\kappa_{\rm eff}\simeq 10^{-10}$, giving rise to an estimated relaxation
rate\cite{weiss,Thorwart04} $\Gamma_{\rm harm}\simeq (\pi/2)\sin^2 (\theta)\, 
\kappa_{\rm eff}\, \epsilon \simeq 10^{-13} \Omega$ (evaluated at low
temperature).  Hence, this illustrates that we can easily achieve the situation 
where $\Gamma_{\rm harm}\ll \gamma$ required for this detection scheme.  Then,
for a waiting time (after which we start the measurement) much longer than the
relaxation time $\gamma^{-1}$ of the nonlinear oscillator, but still smaller 
than $\Gamma^{-1}$, the oscillator is able to reliably detect the qubit state.
\begin{figure}[t!]
\includegraphics*[width=80mm]{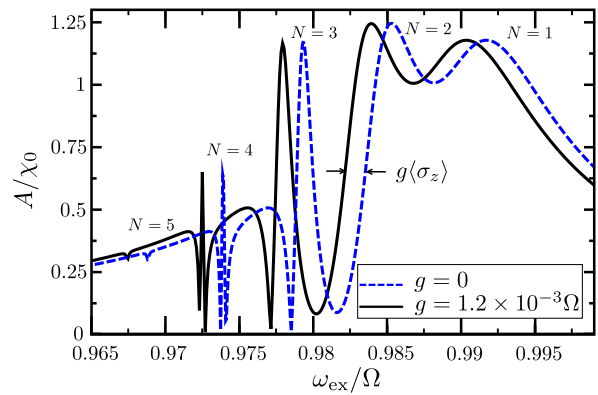}
\caption{(Color online) Nonlinear response $A$ of the detector
         as a function of the external driving frequency $\omega_{\rm ex}$
         in the presence of a finite coupling $g=0.0012 \, \Omega$ to the qubit
         (black solid line). The blue dashed line indicates the response of the
         isolated detector. The parameters are the same as in Fig.\ \ref{fig1}
         and $\epsilon=2.2\Omega$ and $\Delta=0.05\Omega$, in correspondence 
         to realistic experimental parameters \cite{Lupascu05}. \label{fig2}}
\end{figure}
In fact, under these conditions,  the state of the qubit, apart from the
inevitable dephasing, remains unaffected in a time window before it reaches its
global stationary state and an effective shift of the oscillator's
eigenfrequency arises due to the parametric coupling term $\sim g\cos \theta \,
a^\dagger a \, \tau_z$ in Eq.\ (\ref{RWAhamiltonian}).  Treating the
qubit-detector interaction term in Eq.\ (\ref{squidH}) perturbatively to lowest
order in $g$, the eigenfrequency shift follows straightforwardly as 
\[
\Omega \rightarrow \Omega + g \, \langle \sigma_z \rangle \;. 
\]
Thus, the nonlinear response is shifted by $-g$ ($+g$) if the qubit is prepared
in the state $\sigma_z=-1$ ($\sigma_z=1$).  This is illustrated in Fig.
\ref{fig2}, in which we show the nonlinear response of the resonator for the
uncoupled (blue dashed line) and the coupled (black solid line) case. 
\begin{figure}[t!]
\includegraphics[width=80mm]{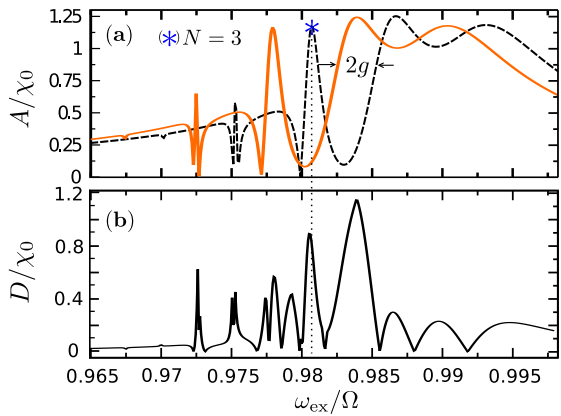}
\caption{(Color online) (a) Nonlinear response $A$ of the detector coupled to
         the qubit prepared in its
         ground state $|\!\!\downarrow\rangle$ (orange solid line) and in its
         excited state $|\!\!\uparrow\rangle$ (black dashed line) for the same
         parameters as in Fig.\ \ref{fig2}.  The quadratic qubit-detector
         coupling induces a global frequency shift of the response by $\delta
         \omega_{\rm  ex}=2 g $. (b) Discrimination power $D(\omega_{\rm ex})$
         of the detector coupled to the qubit for the same parameters as in a).
\label{fig3}}
\end{figure}
For a fixed value of $g$, the shift between the two cases of the opposite qubit
states is given by the frequency gap $\delta \omega_{\rm  ex}\simeq 2\,g$.
Figure \ref{fig3} (a) shows the nonlinear response of the detector for the two
cases
when the qubit is prepared in one of its eigenstates: $|\!\!\uparrow\rangle$
(orange solid line) and $|\!\!\downarrow\rangle$ (black dashed line). 

An important feature of a detection scheme is that it is efficient in
discriminating the states to be detected. This can be quantified by the
discrimination power of the detector, which can be defined for our case as 
\begin{equation}\label{domex}
 D(\omega_{\rm ex}) = \left| 
{A_{|\uparrow \rangle}(\omega_{\rm ex})- A_{|\downarrow
\rangle}(\omega_{\rm ex})}
\right| \, .
\end{equation}
The result for $D(\omega_{\rm ex})$ is shown in Fig.\ \ref{fig3} (b).  The
discrimination power shows a rich structure of local maxima and minima, which
indicates that it can be tuned directly by tuning the driving frequency.  It is
moreover important to realize that the discrimination power can be optimized by
tuning $g$. In the optimized case, a local maximum of the multiphoton resonance
for one qubit state can be made to coincide with a local minimum of the response
for the opposite qubit state yielding to a maximal discrimination power.  An
example where the discrimination power has been optimized with respect to the
three-photon resonance is shown in Fig.\ \ref{fig3} (b).
\section{\label{sec:BackQUBIT} Back action in the qubit}
Another important prerequisite for a useful detection scheme is that the
coupling of the qubit to the detector around a multiphoton resonance does not
generate a destructive back action on the qubit dynamics. In this section, we
show that the back action in this design is surprisingly small for a realistic
choice of parameters.

The back action of the detector on the qubit arises in the form of two
contributions 
from the coupling. First, this coupling has a parametric component ${\cal H}_1 =
g \cos \theta \, n \, \tau_z$, which commutes with the Hamiltonian.  Thus, in
the presence of a coupling of the oscillator to the bath, this term only
produces
dephasing and no relaxation, as it is, for instance, required for a quantum
non-demolition measurement. This part guarantees an efficient detection of the
qubit state.  The second component ${\cal H}_2 = g \sin \theta (a^{\dagger \, 2}
\tau^- + a^2 \tau^+) /2$ in the coupling term yields transitions in the qubit
when two-photon processes are induced in the detector by the external driving
and/or by dissipative transition.  Since, at low temperature, dissipation is
dominated by photon leaking and the driving is very weak, the decay rate of the
qubit from its excited state to its ground state accompanied by the emission of
two oscillators photons,  largely exceeds  the excitation rate  from the ground
state to the excited state accompanied by the absorption of two photons
originally coming from the bath or the driving.  On the other hand, when the
effective oscillator frequency is close to a multiphoton resonance, photon
absorption in the coupled system is enhanced and thus the asymptotic qubit
population might be reduced.   

\begin{figure}[t!]
 \includegraphics[width=80mm]{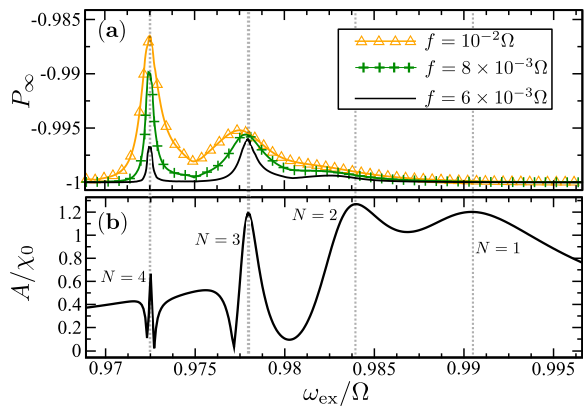}
 \caption{(Color online) (a) Asymptotic population difference $P_\infty$ of the
          qubit states, and (b) the corresponding detector response
          $A$ as a function of the external frequency $\omega_{\rm 
          ex}$ for the same parameters as in Fig.\ \ref{fig2}. \label{fig4}} 
\end{figure}

Thus, for a large asymmetry $|\epsilon|\gg\Delta$, peaks and dips in the qubit
population difference $P_\infty$ are expected  when multiphoton transitions in
the detector are induced.  This is what is shown in Fig. \ref{fig4}(a), where
$P_\infty$ is shown for several values of $f$. For an easier orientation, we
show in addition the corresponding stationary nonlinear response of the detector
in Fig. \ref{fig4}(b).  For increasing driving, the deviation from the expected
value $P_\infty=-1$ becomes more pronounced for larger photon numbers $N$ and
larger driving $f$.  The reason is that, for increasing driving, a
larger Rabi frequency for the corresponding transition results [see Eq.
(\ref{rabi})].  From Fig. \ref{fig4}, it follows that when the qubit is prepared
in its ground state $|\!\!\uparrow\rangle$ (we consider $\epsilon\gg\Delta$)
the back action is very small.  The impact is less than $2 \%$ for the
considered
realistic parameters, yielding to a readout contrast of more than $98 \%$.  This
has to be compared with presently achievable readout contrasts of less than $90
\%$, \cite{Lupascu06} which results from an architecture with a classical 
Josephson bifurcation amplifier.  In passing, we note that the detector response
can also be calculated from the stationary solution of the master equation
(\ref{FME}), but for the parameters considered here (in particular because of
the large qubit bias), this coincides with the shifted one. 

Moreover, we note that the components ${\cal H}_1$ and ${\cal H}_2$ can be 
tuned by $\epsilon$ and $\Delta$. Therefore, $g \sin \theta $ can in principle be 
eliminated by setting $\Delta=0$, which would imply that the measurement scheme 
keeps the state of the qubit without any relaxation but only pure dephasing
(ideal quantum nondemolition measurement). However, turning off the splitting
implies a major change in the experimental design of the sample, since this
parameter is determined by the Josephson energy in the junctions of the
superconducting flux qubit and, thus, may not be easy to be realized. 

The back action of the detector on the qubit should be small not only when the
qubit is in its ground state but also when it is in its excited state. We
therefore address next the relaxation rate of the qubit. Energy relaxation in
the qubit induced by the measurement process will be proportional to the
fluctuations of the square of the phase operator $\chi_+$ induced by the
detector's environment.\cite{Clerk10,Lupascu04}  This relaxation process is
characterized by the transition rate \cite{Clerk10,Lupascu04} 
\begin{equation}\label{relaxrate}
 \Gamma \simeq \tilde{g}^2 \, \sin^2 \theta \, {\cal S}_{\chi_+^2}
[-\omega_{\rm qb}],
\end{equation}
which has been computed perturbatively to lowest order in $\tilde{g}$. Here, 
\begin{eqnarray}
 {\cal S}_{\chi_+^2}[\omega]&=&\frac{\omega_{\rm ex}}{4\pi}
\int_0^{2\pi/\omega_{\rm ex}} \!\!\!\!dt \int_{-\infty}^{+\infty} \!\!\!  d\tau
e^{i\omega \tau} \langle \{
\chi^2_+(\tau+t),\chi^2_+(t)\} \rangle \nonumber \\ &&
\end{eqnarray}
is the symmetrized power spectrum of $\chi_+^2$ averaged over the period of the
external driving (see Appendix for details), with $\{,\}$
indicating the anticommutator. The fact that information on the qubit state is
acquired in the detector via the same channel by which dissipation is introduced
is
nicely reflected in the expression of the relaxation rate in Eq.\
(\ref{relaxrate}). In Fig. \ref{fig5}(a), the relaxation rate $\Gamma$ is shown
for a large negative asymmetry in the qubit. The relaxation rate is strongly
peaked around the multiphoton transitions. There, the noise from the detector
absorbs more energy from the qubit around the multiphoton transition $(0,N)$ 
 since the parametric component ${\cal H}_1$ of the coupling  becomes
negligible, leading to a dominant relaxation process induced by ${\cal H}_2$. 

We emphasize that although the relaxation is maximally enhanced at a multiphoton
resonance, the absolute value of $\Gamma$ is still very small in comparison to
the damping constant, e.g., $\Gamma/\gamma \sim 10^{-6}$. Thus, we can infer the
qubit state with sufficient precision by operating the detector in its steady
state regime as it has been assumed in Section  \ref{nobackaction}. 

\section{efficiency of the measurement}\label{sec:queff}

The measurement of the qubit state requires a coupling to the outer
world, which clearly introduces noise to the qubit. In turn, the noisy detector
yields measurement results, which are statistically distributed. This implies
that several measurements have to be performed to obtain a reliable statistics. 
Hence, the relaxation time of the qubit state should not only exceed the typical
relaxation time of the detector but also the time it takes to acquire sufficient
information to infer the qubit state (the measurement time $T_{\rm  meas}$). 
Hence, for a good measurement fidelity, $T_{\rm meas}$ should be smaller than
the characteristic time $\Gamma^{-1}$ given by Eq. \eqref{relaxrate}, or, 
$\Gamma_{\rm  meas}/\Gamma \gg 1$.  

\begin{figure}[t!]
 \includegraphics*[width=85mm]{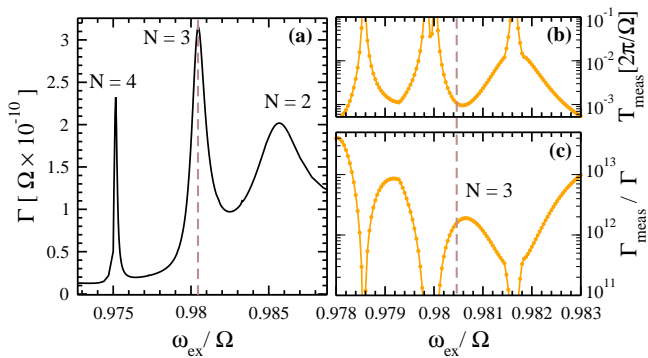}
 \caption{(Color online) (a) Relaxation rate $\Gamma$ of the nonlinear
quantum detector, (b) the measurement time $T_{\rm  meas}$,  and (c) 
the measurement efficiency $\Gamma_{\rm  meas}/\Gamma$
as a function of the 
          external frequency $\omega_{\rm  ex}$. The parameters are
          the same as in Fig.\ \ref{fig2}. \label{fig5}} 
\end{figure}

The measurement time can be formalized \cite{Lupascu04,Clerk10,Makhlin01} as the
ratio of the symmetrized power spectrum ${\cal S}_{\chi_+}$ of the phase
operator $\chi_+$ (evaluated at zero frequency) and the square of the difference
between the two expectation values of $\chi_+$ when the qubit is in the two
opposite states, i.e., with Eq.\ (\ref{domex}),  
\begin{equation}
 T_{\rm  meas} = \frac{{\cal S}_{\chi_+} }{[D(\omega_{\rm ex})]^2}
\,\, .
\end{equation}
The result for $T_{\rm  meas}$ as a function of $\omega_{\rm ex}$  is shown in
Fig.\ \ref{fig5} b) for the parameter set used above, for which the
discrimination power  $D(\omega_{\rm ex})$ around the 3-photon resonance has
been maximized. In correspondence with this is the relative minimum 
of $T_{\rm  meas}$ around the 3-photon resonance, see Fig.\ \ref{fig5} b).
Interestingly enough, 
the time scale of the measurement time around this resonance 
is $T_{\rm  meas}\approx 10^{-2}\times 2\pi /\Omega$. Considering 
realistic numbers of a typical experimental set-up \cite{Lupascu05}, where 
  $\Omega$ is in the regime of a few GHz, we obtain a time scale of 
 $T_{\rm  meas}\approx100 $ ps for the nonlinear quantum detection
scheme. This should be contrasted to the measurement time of $T_{\rm 
meas}\approx 300 $ ns obtained in Ref.\ \onlinecite{Lupascu05}. 
In between the multiphoton resonances, the dependence of $T_{\rm  meas}$ on 
$\omega_{\rm ex}$ shows a rich structure including several singularities, which
are simply due to the several crossings of the two nonlinear response curves
shown in Fig.\ \ref{fig3} a), where $D(\omega_{\rm ex})$ becomes zero, implying
insufficient discrimination of the two qubit states. 

 With this, we can evaluate the measurement
efficiency, defined by the ratio $\Gamma_{\rm  meas} / \Gamma$, with
$\Gamma_{\rm  meas}=T_{\rm  meas}^{-1}$. This quantity sets the probability
to infer the qubit state, based on the nonlinear response of the detector. We
show the result for the efficiency of the measurement in Fig.\  \ref{fig5}(c).
Related to the multiphoton resonances in the detector, the efficiency also shows
local maxima. For the discrimination power being optimized around the
three-photon resonance, the measurement efficiency displays a clear 
local maximum [see Fig. \ref{fig5}(c)]. Due to the small size of the relaxation
rate $\Gamma$ of the detector, the overall measurement efficiency is rather
large in comparison  to the detection set-up with a linear resonator,
\cite{Lupascu05} ensuring $\Gamma_{\rm  meas}/\Gamma \gg 1$.  
\section{Conclusions}
To conclude, we have introduced a scheme for quantum state detection on
the basis of a nonlinear detector which is operated in the regime of
resonant few-photon transitions. Discrete multiphoton resonances in the
detector can be used to infer
the state of the parametrically coupled qubit via a state-dependent frequency
shift of the
detector's nonlinear response function. The multiphoton resonances are well
separated in the spectrum and sharp enough to allow for a
good resolution of the
qubit state. 

By analyzing key quantities of the detector, we have shown that the nonlinear
few-photon detector can be operated efficiently, reliably, and with sufficiently
weak back action. In fact, it can be efficiently tuned
by tuning the amplitude of the ac bias current of the SQUID. Furthermore, we
have shown that the sharpness of the multiphoton resonances can be used to
obtain an increased discrimination power as compared to the linear parametric
detection scheme. Clearly, the relaxation rate at a multiphoton
resonance for the qubit becomes maximal, but in general remains very small.
The measurement time around a multiphoton resonance can be tuned such that it
becomes minimal. For realistic experimental parameters, we find surprisingly
small measurement times, allowing in principle for fast measurements. Moreover, 
the efficiency of the measurement, which takes the time to acquire
enough information to infer the qubit state into account, also assumes large
values, thus allowing for a reliable and highly efficient measurement of the
qubit state. 

We have chosen realistic values for the involved model parameters such
that an experimental realization of this quantum measurement scheme should
become possible in the near future. The nonlinear detection scheme in the deep
few-photon quantum regime offers thus the advantage of an increased
discrimination power of more than $98 \%$ (for our choice of realistic
parameters), as compared to previous classical detection schemes based 
on the Josephson bifurcation amplifier. 

A possible setup in order to realize the nonlinear few-photon detector could
be the architecture used in a recent experiment.\cite{Murch10} 
The low-temperature regime, where quantum noise effects are important, has
already been reached. In order to operate in the regime of only few photons 
in the resonator, the sensitivity and stability of the devices might have 
still to be further increased. However, no principle obstacles are apparent.
%
%
\begin{acknowledgements}

This work was supported by the DAAD (German Academic Research Service) 
Research Grant No. Ref: A/08/73659. V. P. was supported by the NSF (Grant
No. EMT/QIS 082985).  We thank P. Nalbach for valuable discussions.  

\end{acknowledgements}


\appendix*

\section{\label{App} Power spectrum}

To calculate the power spectrum of a driven out-off-equilibrium system is a
non trivial task, since the time reversal symmetry, which simplifies the
calculation in equilibrium systems, is not given anymore. An elegant way to
compute this in terms of correlation functions is presented in Ref.
\onlinecite{Melvin63}. The correlation function between two operators $A$ and
$B$ is described as a mean value
\begin{equation}
 S_{\rm  AB}(t,\tau) = {\rm Tr}_{\rm  S \oplus B} \lbrace {\cal W}(t + \tau) 
A(0) \rbrace .
\end{equation}
Here, the trace is over the whole system-plus-bath, with the ``density''
operator ${\cal W} (t+\tau) =  U(t+\tau,t) [B W (t)] U^\dagger (t+\tau,t)$ and
$U(t+\tau,t) = \exp \lbrace  - i {\cal T} \int_{t}^{t+\tau} H_{\rm total} (t') d
t' \rbrace $, with $H_{\rm  total}$ being the Hamiltonian of the whole
system-plus-bath, $W$ the density operator of the total system, and ${\cal T}$
the time ordering operator. Furthermore, $A$ and $B$ are in the Heisenberg
representation. 

In the regime of weak coupling to the environment, the reduced density operator
$\tilde{\varrho} (t+\tau) \equiv {\rm Tr}_B \lbrace {\cal W} (t+\tau) \rbrace$
evolves according to the master equation (\ref{FME}).

In the superoperator notation, ${\cal D}_{\alpha \beta, \alpha' \beta'}
= -i(\varepsilon_\alpha - \varepsilon_\beta)\delta_{\alpha \alpha'} 
\delta_{\beta \beta'} + {\cal L}_{\alpha \beta, \alpha' \beta'}$ (Liouville
superoperator) is represented by a ${\cal N}^2 \times {\cal N}^2$ supermatrix
${\bf \cal D}$, where ${\cal N}$ is the number of effective states in the
system. In the same way, the density operator $\varrho_{\alpha \beta}$ formally
is a ${\cal N}^2$ dimensional column vector $\varrho (t)$. The solution of the
master equation is reduced to an eigenvalue problem of the matrix ${\bf \cal
D}$, as 
\begin{eqnarray}
  {\bf \cal D} \cdot {\bf v}^m 
= \Gamma_m {\bf v}^m ,  \quad  {\bf v}^\dagger_m \cdot {\bf \cal D} 
= \Gamma_m {\bf v}^\dagger_m , 
\end{eqnarray}
where ${\bf v}^m$ and ${\bf v}_m$ are the left and right eigenvectors,
respectively, with eigenvalue $\Gamma_m$. 

In the superoperator notation, the master equation \eqref{FME} is expressed as 
$\dot{\varrho}(t) = {\bf \cal D} \cdot{ \varrho}(t)$, and its solution is given
by 
\begin{equation}
 \varrho (t)  =  \exp \lbrace {\bf \cal D }\, t \rbrace \cdot \varrho(t=t_0).
\end{equation}
In the regime of the RWA, and at low temperature the master equation
\eqref{FME} conserves the trace and the positivity of the density operator,
i.e., it assumes  Lindblad form. Therefore, we can expand the solution of the
master equation in terms of the right-eigenvector ${\bf v}^m$
\begin{equation}
 \varrho (t)  =  \sum_m  {\bf v}^m  \,c_m  \exp \lbrace {\Gamma_m \, t
} \rbrace, 
\end{equation}
with $c_m = {\bf v}^\dagger_m \cdot \varrho(t_0)$. Here, we have used the
orthogonality property ${\bf v}_m^\dagger \cdot {\bf v}^{m'} = \delta_{mm'}$.  
The corresponding expression follows for the operator $\tilde{\varrho}(t+\tau)$,
but with a different initial condition. It is easily understandable in the
operator notation  
\begin{equation}
\tilde{\varrho}_{\alpha \beta}(t + \tau) = \sum_{m,\alpha'\beta'\xi'}
v^m_{\alpha \beta} \; v^\dagger_{m,\beta'\alpha'} \; 
B_{\alpha' \xi'} (t) \; \varrho_{\xi' \beta'}(t)\;e^{\Gamma_m \tau}
\end{equation}
where $B_{\alpha \beta} (t) = \langle \phi_{\alpha} (t)| B |\phi_{\beta} (t)
\rangle$, and $v^m_{\alpha \beta}$ ($v_{m,\alpha \beta}$) is the operator
representation of ${\bf v}^m$ (${\bf v}_m$) in the quasienergy states.
Considering an initial time in the stationary regime, i.e., $\varrho_{\alpha
\beta} (t) \rightarrow \varrho^\infty_{\alpha \beta}$, and after averaging the
initial time $t$ over the period $2 \pi/\omega_{\rm  ex}$, the correlation
function reads
\begin{equation} \label{CorrF}
 S_{\rm  AB}(t) = \sum_{m,n} \; S_{nm} \; e^{\Gamma_m t - i n \omega_{\rm  ex}
t} , 
\end{equation}
with
\begin{equation}
 S_{nm} =  \sum_{\alpha \beta} \sum_{\alpha' \beta' \xi' }
\; v^m_{\alpha \beta} \; A_{\beta \alpha, n} \; v^\dagger_{m,\beta'\alpha'} \;
B_{\alpha' \xi',-n}\; \rho^\infty_{\xi'\beta'}, 
\end{equation}
where $A_{\alpha \beta,n}$ and $B_{\alpha \beta ,n}$ are the coefficients of
the Fourier expansion of $A_{\alpha \beta}(t)$ and $B_{\alpha \beta}(t)$,
respectively. The power spectrum is obtained directly from the Fourier
transform of Eq. (\ref{CorrF}).

\end{document}